\documentclass[12pt]{article}
\usepackage{graphics,epsfig}
\usepackage[english]{babel}
\newcommand{\cinst}[2]{$^{\mathrm{#1}}$~#2\par}
\newcommand{\crefi}[1]{$^{\mathrm{#1}}$}

\begin{document}

%\begin{titlepage}

\thispagestyle{empty}
%%%%%%%%%%%%%%%%%%%%%%%% COVER PAGE
\begingroup

%\raisebox{0.5cm}[0cm][0cm] {
%\begin{tabular*}{\hsize}{@{\hspace*{5mm}}ll@{\extracolsep{\fill}}r@{}}
%\begin{minipage}[t]{3cm}
%\vglue.5cm
%\end{minipage}
%&
%\begin{minipage}[t]{7cm}

%\end{minipage}
%&
%\begin{minipage}[t]{7cm}
%\vglue.5cm {\bf YerPhI Preprint 1582(3)-2003} %\vglue.1cm

%\end{minipage}
%\end{tabular*}
%}

\begin{center}
%{\large{YEREVAN PHYSICS INSTITUTE}}

%\hspace{9cm}{Draft 1}
\vglue 1.0cm {\Large{\bf The A- dependence of $\rho^0$
neutrinoproduction on nuclei}}

%(DRAFT 1)
\end{center}

\vspace{1.cm}

\begin{center}
{\large SKAT Collaboration}

 N.M.~Agababyan\crefi{1}, V.V.~Ammosov\crefi{2},
 M.~Atayan\crefi{3},\\
 N.~Grigoryan\crefi{3}, H.~Gulkanyan\crefi{3},
 A.A.~Ivanilov\crefi{2},\\ Zh.~Karamyan\crefi{3},
V.A.~Korotkov\crefi{2}

\setlength{\parskip}{0mm}
\small
%\HRule\\

\vspace{1.cm} \cinst{1}{Joint Institute for Nuclear Research,
Dubna, Russia} \cinst{2}{Institute for High Energy Physics,
Protvino, Russia} \cinst{3}{Yerevan Physics Institute, Armenia}
\end{center}
\vspace{100mm}

{\centerline{\bf YEREVAN  2005}}

%\end{titlepage}

\newpage
\vspace{1.cm}
\begin{abstract}
The $A$- dependence of $\rho^0$ meson production in
neutrino-induced reactions is investigated for the first time,
using the data obtained with SKAT bubble chamber. The nuclear
medium influence on the $\rho^0$ total yield and inclusive
distributions (on $z = E_{\rho}/\nu$ and Feynman $x_F$ variables)
is found to be approximately the same as for pions. It is shown,
that these distributions, with increasing $A$, tend to shift
toward smaller values of $z$ and $x_F$, thus indicating on an
increasing role of secondary intranuclear interactions. The
predictions of a simplified model incorporating the latter are
found to be in qualitative agreement with experimental data.
\end{abstract}

\newpage
\setcounter{page}{1}
\section{Introduction}

%\vglue2cm
The total yields and inclusive spectra of different species of
hadrons in leptonuclear interactions reflect the complicated
space-time structure of the quark string fragmentation, the hadron
formation and the secondary intranuclear interactions. The
experimental investigations on this topic concern mainly stable
hadrons (pions, kaons, (anti)protons). Meanwhile, the data
concerning hadronic resonances (being, predominantly, direct
products of the quark string fragmentation) can provide a valuable
additional information about the nuclear medium influence on the
leptoproduction processes. Hitherto, no data are available for the
A- dependence of the total yields and inclusive spectra of
hadronic resonances in lepton-induced reactions. \\ This work is
devoted to the study, for the first time, of the A- dependence of
$\rho^0$ meson neutrinoproduction on nuclei. To this end, the data
from the SKAT bubble chamber are used. In Section 2, the
experimental procedure is described. The experimental data on the
A- dependence of the total yield and inclusive spectra of $\rho^0$
are presented in Section 3 and discussed in Section 4. The results
are summarized in Section 5.

\section{Experimental procedure}

\noindent The experiment was performed with SKAT bubble chamber
\cite{ref1}, exposed to a wideband neutrino beam obtained with a
70 GeV primary protons from the Serpukhov accelerator. The chamber
was filled with a propane-freon mixture containing 87 vol\%
propane ($C_3H_8$) and 13 vol\% freon ($CF_3Br$) with the
percentage of nuclei H:C:F:Br = 67.9:26.8:4.0:1.3 \%. A 20 kG
uniform magnetic field was provided within the operating chamber
volume.
\\ Charged current interactions containing a negative muon with momentum
$p_{\mu} >$0.5 GeV/c were selected. Other negatively charged
particles were considered to be $\pi^-$ mesons. Protons with
momentum below 0.6 GeV$/c$ and a fraction of protons  with
momentum 0.6-0.85 GeV$/c$ were identified by their stopping in the
chamber. Non-identified positively charged particles were
considered to be ${\pi}^+$ mesons. Events in which errors in
measuring the momenta of all charged secondaries and photons were
less than 60\% and 100\%, respectively, were selected. Each event
is given a weight which corrects for the fraction of events
excluded due to improperly reconstruction. More details concerning
the experimental procedure, in particular, the reconstruction of
the neutrino energy $E_{\nu}$ can be found in our previous
publications \cite{ref2,ref3}. \\ The events with $3 < E_{\nu} <$
30 GeV were accepted, provided that the reconstructed mass $W$ of
the hadronic system exceeds 2 GeV. No restriction was imposed on
the transfer momentum squared $Q^2$. The number of accepted events
was 4353 (5508 weighted events). The mean values of the
kinematical variables were $<E_{\nu}>$ = 10.7 GeV, $<W>$ = 3.0
GeV, $<W^2>$ = 9.6 GeV$^2$, $<Q^2>$ = 2.8 (GeV/c)$^2$. \\ Further,
the whole event sample was subdivided, using several topological
and kinematical criteria \cite{ref3,ref4}, into three subsamples:
the 'cascade' subsample $B_S$ with a sign of intranuclear
secondary interaction, the 'quasiproton' ($B_p$) and
'quasineutron' ($B_n$) subsamples. About 40\% of subsample $B_p$
is contributed by interactions with free hydrogen. Weighting the
'quasiproton' events with a factor of 0.6, one can compose a
'pure' nuclear subsample $B_A = B_S + B_n + 0.6 B_p$ and a
'quasinucleon' subsample $B_N = B_n + 0.6 B_p$. It has been
verified \cite{ref4,ref5}, that the multiplicity and spectral
characteristics of secondary particles in the $B_p (B_N)$
subsample are in satisfactory agreement with those measured with a
pure proton (deuteron) target. The effective atomic weight
corresponding to the subsample $B_A$ is estimated \cite{ref6} to
be approximately equal to $A_{eff} = 21 \pm 2$, when taking into
account the probability of secondary intranuclear interactions in
the composite target. \\ In order to extract the A- dependence of
the $\rho^0$ neutrinoproduction, we use in the next section the
data obtained for subsamples $B_N$ and $B_A$, as well as the
published data \cite{ref7} on neutrino-freon interactions (with
$A_{eff} = 45 \pm 2$).

\section{The A- dependence of the $\rho^0$ mean multiplicity and
inclusive spectra}

The $(\pi^+ \pi^-)$ effective mass distribution for subsamples
$B_N$ and $B_A$ is plotted in Fig. 1. Clear signals near the
$\rho^0$ mass, as well as faint signals near the $f_0(980)$ mass
are seen. The distributions are fitted by expression

\begin{equation}
dN/dM_{\pi^+\pi^-} = BG \cdot (1 + \alpha_{\rho} \cdot BW_{\rho} +
\alpha_{f} \cdot BW_{f}) \, ,
\end{equation}

\noindent where the mass dependence of the background distribution
is parametrized as

\begin{equation}
BG = \alpha_1 \cdot exp (\alpha_2 \cdot M + \alpha_3 \cdot M^2) \,
,
\end{equation}

\noindent while for $BW_{\rho}$ and $BW_f$ the corresponding
relativistic Breit-Wigner functions \cite{ref8} for $\rho^0$ and
$f_0(980)$ were used, taking into account the experimental mass
resolution $\sigma(M)$ = 35 MeV. The mass and width of resonances
are fixed as: $M_{\rho}$ = 775 MeV and $\Gamma_{\rho}$ = 150 MeV
from the PDG data \cite{ref9} and $M_f$ = 963 MeV and $\Gamma_f$ =
35 MeV from the recent NOMAD measurements \cite{ref10}. \\ The
resulting total yields of $\rho^0$ and $f_0(980)$ are presented in
Table 1, where the SKAT data for $\rho^0$ in neutrino-freon
interactions \cite{ref7} are also given. The data on $f_0(980)$
are corrected for the undetectable mode $(\pi^0\pi^0)$. For
comparison, the data on the $\pi^-$ yields are shown too.

\begin{table}[ht]
\caption{The mean multiplicities of $\rho^0$, $f_0(980)$ and
$\pi^-$ and the ratio $<n_{\rho^0}>/<n_{\pi^-}>$.}
\begin{center}
\begin{tabular}{|l|c|c|c|c|}
  \hline
%\multicolumn{5}{|c|}{} \\
% &\multicolumn{4}{c|}{4 $< W^2 <$ 25
%GeV$^2$}\\  \multicolumn{5}{|c|}{} \\ $h^+(x_F > 0)$
$A$&$<n_{\rho^0}>$&$<n_{f_0}(980)>$&$<n_{\pi^-}>$&
$<n_{\rho^0}>/<n_{\pi^-}>$
\\ \hline
1&0.070$\pm$0.031&0.030$\pm$0.015&0.73$\pm$0.02 &0.096$\pm$0.043
\\
21&0.075$\pm$0.023&0.019$\pm$0.011&0.80$\pm$0.01 &0.094$\pm$0.029
\\
45&0.09$\pm$0.02&$-$&0.90$\pm$0.01 &0.10$\pm$0.02
\\ \hline

\end{tabular}
\end{center}
\end{table}

\noindent The A-dependences of the $\rho^0$ and $\pi^-$ yields,
plotted in Fig. 2, are rather similar. An exponential
parametrization ($\sim A^{\beta}$) of the yields leads to
compatible values of ${\beta}_{\rho^0}$ = 0.07 $\pm$ 0.13 and
${\beta}_{\pi^-}$ = 0.052 $\pm$ 0.007. As a result, no A-
dependence of the $<\rho^0/\pi^->$ ratio is observed (the last
column of Table 1). \\  Note, that a comparison with higher-energy
neutrino-induced data shows that this ratio tends to increase with
increasing $W$, reaching up to 0.128 $\pm$ 0.030 at $<W>$ = 4.8
GeV \cite{ref11} and 0.156 $\pm$ 0.028 at $<W>$ = 6.1 GeV
\cite{ref12}. \\ Our estimations of the $f_0(980)$ total yield are
rather rough. They do not contradict, within the experimental
uncertainties, the only published data \cite{ref10},
$<n_{f_0}(980)>$ = 0.018 $\pm$0.004, obtained at higher neutrino
energies ($<E_{\nu}>$ = 45 GeV).

\begin{table}[ht]
\caption{The yields of $\rho^0$ and $\pi^-$ and their ratio in the
forward ($x_F > 0$) and backward ($x_F < 0$) hemispheres.}
\begin{center}
\begin{tabular}{|l|ccc|}
  \hline
%\multicolumn{5}{|c|}{} \\
% &\multicolumn{4}{c|}{4 $< W^2 <$ 25
%GeV$^2$}\\  \multicolumn{5}{|c|}{} \\ $h^+(x_F > 0)$
$A_{eff}$&$<n_{\rho^0}>$&$<n_{\pi^-}>$& $<n_{\rho^0}>/<n_{\pi^-}>$
\\ \hline
&& $x_F >0$&  \\ 1&0.066$\pm$0.026&0.435$\pm$0.014&0.15$\pm$0.06
\\
21&0.050$\pm$0.016&0.399$\pm$0.009&0.12$\pm$0.04
\\ \hline
&& $x_F <0$&  \\ 1&0.005$\pm$0.017&0.239$\pm$0.012 &0.02$\pm$0.06
\\
21&0.025$\pm$0.016&0.402$\pm$0.009 &0.06$\pm$0.04 \\ \hline
\end{tabular}
\end{center}

\end{table}

\noindent Table 2 presents the yields of $\rho^0$ and $\pi^-$ and
their ratio in the forward ($x_F > 0$, $x_F$ being the Feynman
variable) and backward ($x_F < 0$) hemispheres in the hadronic
c.m.s. The data on $\pi^-$ indicate, in accordance with our
previous studies (see details in \cite{ref3,ref5,ref6}) a clear
depletion in the forward hemisphere (due to the nuclear
attenuation) and a significant enhancement in the backward
hemisphere (due to the secondary inelastic intranuclear
interactions). A faint indication on similar effects are also seen
for $\rho^0$.  \\ As it is seen from Table 2 and Fig. 3 where the
distributions on $x_F$ for $\rho^0$ and $\pi^-$ are plotted, the
overwhelming part of $\rho^0$ in $\nu N$ interactions and the most
part of that in nuclear interactions are produced in the forward
hemisphere, as a result of the current quark ($u$ or $\bar{d}$)
fragmentation into a favorable hadron $\rho^0$ (which can contain
the current quark), while the production of the (unfavorable)
$\pi^-$ meson occurs mainly in the central region. As a result,
the ratio $<\rho^0/\pi^->$ for the fastest hadrons (with $x_F >
0.5$) significantly exceeds that in other ranges of the variable
$x_F$, as it can be seen from Fig. 4. Note, that a comparison of
our data with those obtained at higher energies does not reveal
any $W$- dependence of the $<\rho^0/\pi^->$ ratio in the forward
hemisphere. Indeed, the values of ${<\rho^0/\pi^->}_N$ =
2.42$\pm$1.08 at $x_F >$ 0.5 for $\nu N$ - interactions and
${<\rho^0/\pi^->}_A$ = 0.12$\pm$0.04 at $x_F >$ 0 for $\nu A$ -
interactions obtained in this work are compatible, respectively,
with ${<\rho^0/\pi^->}_p$ = 1.99$\pm$0.44 measured in $\nu p$ -
interactions at $<W>$ = 6.1 GeV \cite{ref12} and
${<\rho^0/\pi^->}_{Ne}$ = 0.13$\pm$0.03 measured in $\nu {Ne}$ -
interactions at $<W>$ = 4.8 GeV \cite{ref11}. \\ It is interesting
to compare the relative yield of strange and non-strange favorable
mesons, $K^+(890)$ and $\rho^0$, with that of unfavorable ones,
$K^0$ and $\pi^-$. The former can be extracted using our recent
data on $K^+(890)$ neutrinoproduction \cite{ref13}. The ratio of
the $K^+(890)$ and $\rho^0$ total yields is estimated to be
0.19$\pm$0.14 (for $A$ = 1) and 0.20$\pm$0.11 (for $A \approx$ 21)
which seems to exceed that for $K^0$ and $\pi^-$ measured in
\cite{ref6}: 0.055$\pm$0.013 (for $A$ = 1) and 0.070$\pm$0.011
(for $A \approx$ 21). These data, therefore, indicate, that the
strangeness content in favorable mesons in neutrino-induced
reactions is higher than that for unfavorable ones.

\noindent In Figs. 5 - 7, the inclusive spectra of $\rho^0$ for
'quasinucleon' and nuclear interactions are compared with those
measured in neutrino-freon interactions \cite{ref7}. \\ Fig. 5
shows the invariant distribution on $x_F$. It is seen, that the
data do not exhibit a significant $A$- dependence in the forward
hemisphere, while at $x_F <$ 0 the $\rho^0$ yield in $\nu A$ -
interactions is enhanced as compared to $\nu N$ - interactions.
\\ The distributions on the variable $z = E_{\rho^0}/\nu$ are
presented in Fig. 6. It is seen, that they tend to shift towards
lower values of $z$ with increasing A, as expected due to the
effects of secondary intranuclear interactions. Fig. 7 shows the
$z$- dependence of the ratio $\rho^0/({\pi^+} +\pi^-)/2$. The data
exhibit no significant dependence on $A$, thus indicating that the
nuclear effects are approximately the same for $\rho^0$ and pions
(averaged over $\pi^+$ and $\pi^-$).

\section{Discussion}

The data presented in the previous section indicate on a small but
not negligible role of nuclear effects in the $\rho^0$
neutrinoproduction. It is interesting to clarify whether these
effects are compatible with expectations based on the accounting
for intranuclear interactions of secondary pions, $\pi N
\rightarrow \rho^0 X$, resulting in a $\rho^0$ multiplicity gain
$\delta_{\rho^0}^{sec}$ (mainly at $x_F < 0$), as well as the
absorption processes, $\rho^0 N \rightarrow$ (no $\rho^0$),
resulting in a yield reduction, $\delta_{\rho^0}^{abs}$ (mainly at
$x_F > 0$). The details of a simple model incorporating these
processes can be found in \cite{ref6} and references therein. \\
The gain $\delta_{\rho^0}^{sec}(p_{\pi})$ induced by pions with
momenta $p_{\pi} \pm {\Delta}p_{\pi}$ is determined by the
differential multiplicity of pions, ${\Delta}n(p_{\pi})$, the mean
probability of their secondary inelastic interactions
$<w_A(p_{\pi})>$ averaged over the nuclei of the composite target,
and by the mean multiplicity $\bar{n}_{\rho^0}(p_{\pi})$ of
$\rho^0$ in inelastic $\pi N$ interactions. The values of
$\bar{n}_{\rho^0}(p_{\pi})$ extracted (with an uncertainty of
about $\pm$ 15\%) from the available experimental data
\cite{ref14} vary from 0.002 at $p_{\pi}$ = 0.9 - 1 GeV$/c$ up to
0.22 at $p_{\pi}$ = 10 - 15 GeV$/c$ (the end of the pion spectrum
in this experiment). \\  The product
$\delta_{\rho^0}^{sec}(p_{\pi}) = {\Delta}n(p_{\pi}) \cdot
<w_A(p_{\pi})> \cdot \, \bar{n}_{\rho^0}(p_{\pi})$ was integrated
over the momentum spectra of charged pions measured in
'quasinucleon' interactions. The contribution from secondary
interactions of $\pi^0$ mesons is assumed to be the average of
those for $\pi^+$ and $\pi^-$ mesons. The resulting values of
${\delta}^{sec}_{\rho^0}$ are found to be 0.031$\pm$0.005 for
$A_{eff} \approx$ 45 and 0.023$\pm0.003$ for $A_{eff} \approx$ 21.
The latter value can be compared with the experimental value of
${\delta}^{exp}_{\rho^0}(x_F < 0)$ = 0.020$\pm$0.012, inferred
from the last two lines of Table 2 (note, that in the evolution of
the error in ${\delta}^{exp}_{\rho^0} = {<n_{\rho^0}>}_A -
{<n_{\rho^0}>}_N$ the correlation between the values of
${<n_{\rho^0}>}_A$ and ${<n_{\rho^0}>}_N$ was taken into account
here and below). \\ In order to estimate the $\rho^0$ yield
decreasing, ${\delta}^{abs}_{\rho^0}$, one needs to know the
$\rho^0$ absorption cross section ${\sigma}^{abs}_{\rho^0}$ via
the channel $\rho^0 N \rightarrow$ (no $\rho^0$). As
${\sigma}^{abs}_{\rho^0}$ is not known, we use tentative values in
between $5 < {\sigma}^{abs}_{\rho^0} < 10$ mb. At these values,
the probability $w^{abs}_{\rho^0}$ of the $\rho^0$ absorption is
estimated to be 23$\pm$6\% for $A_{eff} \approx$ 45 and 18$\pm$5\%
for $A_{eff} \approx$ 21 (see \cite{ref3} for details of the
absorption probability calculations). With these probabilities,
the value of ${\delta}^{abs}_{\rho^0} = -w^{abs}_{\rho^0} \cdot \,
{<n_{\rho^0}(x_F > 0)>}_N$ is equal to -0.016$\pm$0.008 and
-0.012$\pm$0.006, respectively. The latter value can be compared
with ${\delta}^{exp}_{\rho^0}(x_F > 0)$ = -0.016$\pm$0.020
inferred from the first two lines of Table 2. \\ The total
multiplicity gain, ${\delta}_{\rho^0} =
{\delta}^{sec}_{\rho^0}(x_F < 0) + {\delta}^{abs}_{\rho^0}(x_F > 0
)$, is predicted to be  ${\delta}_{\rho^0}$ = 0.011$\pm$0.007 for
$A_{eff} \approx$ 21 and 0.015$\pm$0.009 for $A_{eff} \approx$ 45.
These values are compatible with a rather weak variation of
$<n_{\rho^0}>$ with $A$ (cf. Table 1), resulting in a total
multiplicity gain ${\delta}^{exp}_{\rho^0}$ = 0.005$\pm$0.020 for
$A_{eff} \approx$ 21 and 0.020$\pm$0.037 for $A_{eff} \approx$ 45.
The large errors in ${\delta}^{exp}_{\rho^0}$ do not allow to
check quantitatively the predictions of the model incorporating
secondary intranuclear interactions.

\section{Summary}

New experimental data on $\rho^0$ production in $\nu N$ and $\nu
A$ ($A \approx 21$) interactions are obtained at intermediate
energies ($<E_{\nu}>$ = 10.7 GeV, $<W>$ = 3.0 GeV). For the first
time, nuclear effects in $\rho^0$ neutrinoproductions are
observed. These effects for the total yield $<n_{\rho^0}>$ of
$\rho^0$ are found to be rather small and compatible with those
for pions. Using also the SKAT data for $A_{eff} \approx$ 45, the
slope parameter $\beta$ in the exponential parametrization of the
total yields ($<n> \sim A^{\beta}$) is found $\beta_{\rho^0} =
0.07 \pm 0.13$ which agrees with that for $\pi^-$ mesons,
$\beta_{\pi^-} = 0.052 \pm 0.007$. The ratio of $<n_{\rho^0}> /
<n_{\pi^-}>$ is found to be independent of $A$ being equal about
0.1. A comparison with the data at higher energies reveals a
slight increase of this ratio with increasing energy, while no
energy dependence is observed for that in the forward hemisphere
($x_F > 0$). \\ A comparison of the $\rho^0$ inclusive spectra in
$\nu N$ and $\nu A$- interactions indicates, that the major
influence of the nuclear medium consists in their shifting towards
smaller values of $x_F$ and $z$, thus indicating on a
non-negligible role of the secondary intranuclear interactions.
The predictions of a simplified model incorporating the latter are
found to be in qualitative agreement with experimental data. \\ A
comparison of the relative yield of the favorable strange and
non-strange mesons ($K^+(890)$ and $\rho^0$) with that for
unfavorable ones ($K^0$ and $\pi^-$) indicates, that the
strangeness content in the former is higher than in the latter. \\
{\bf Acknowledgement.} The activity of one of the autors (H.G.) is
supported by Cooperation Agreement between DESY and YerPhI signed
on December 6, 2002. The autors from YerPhI acknowledge the
supporting grants of Calouste Gulbenkian Foundation and Swiss
Fonds "Kidagan".

%%%%%%%%%%%%%%%%%%%%%%%%%%%%%%%%%%%%%%%%%%%%%%%%%%%%%%%%%%
   %%% References
%%%%%%%%%%%%%%%%%%%%%%%%%%%%%%%%%%%%%%%%%%%%%%%%%%%%%%%%%%

\newpage
\begin{figure}[ht]
 \resizebox{0.9\textwidth}{!}{\includegraphics*[bb =20 160 600
510]{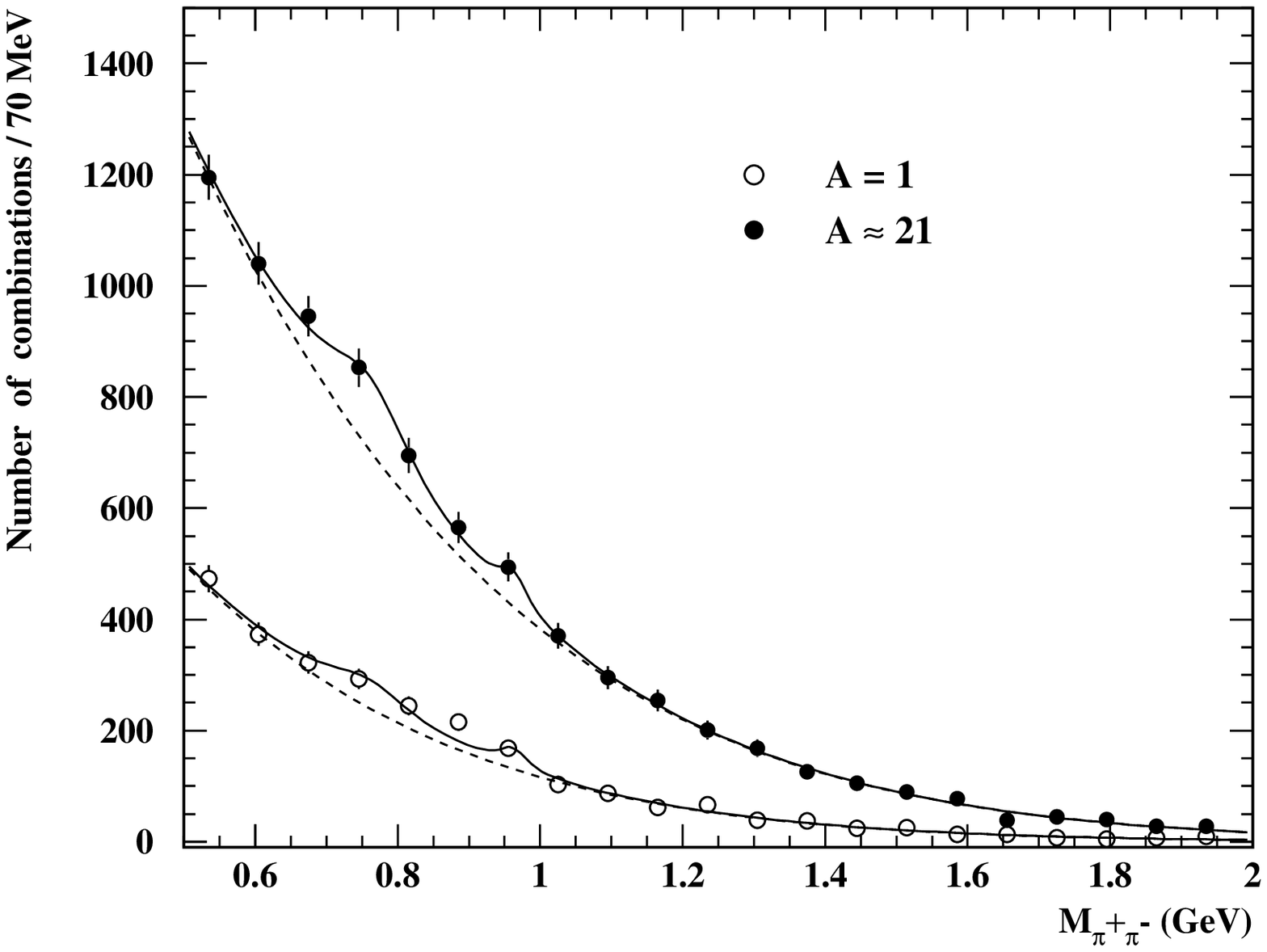}} \caption{The effective mass distribution for the
system ($\pi^+ \pi^-$). The curves are the result of the fit (see
text).}
%\end{figure}

%\newpage
%\begin{figure}[h]
\resizebox{0.7 \textwidth}{!}{\includegraphics*[bb=20 160 500 550]
{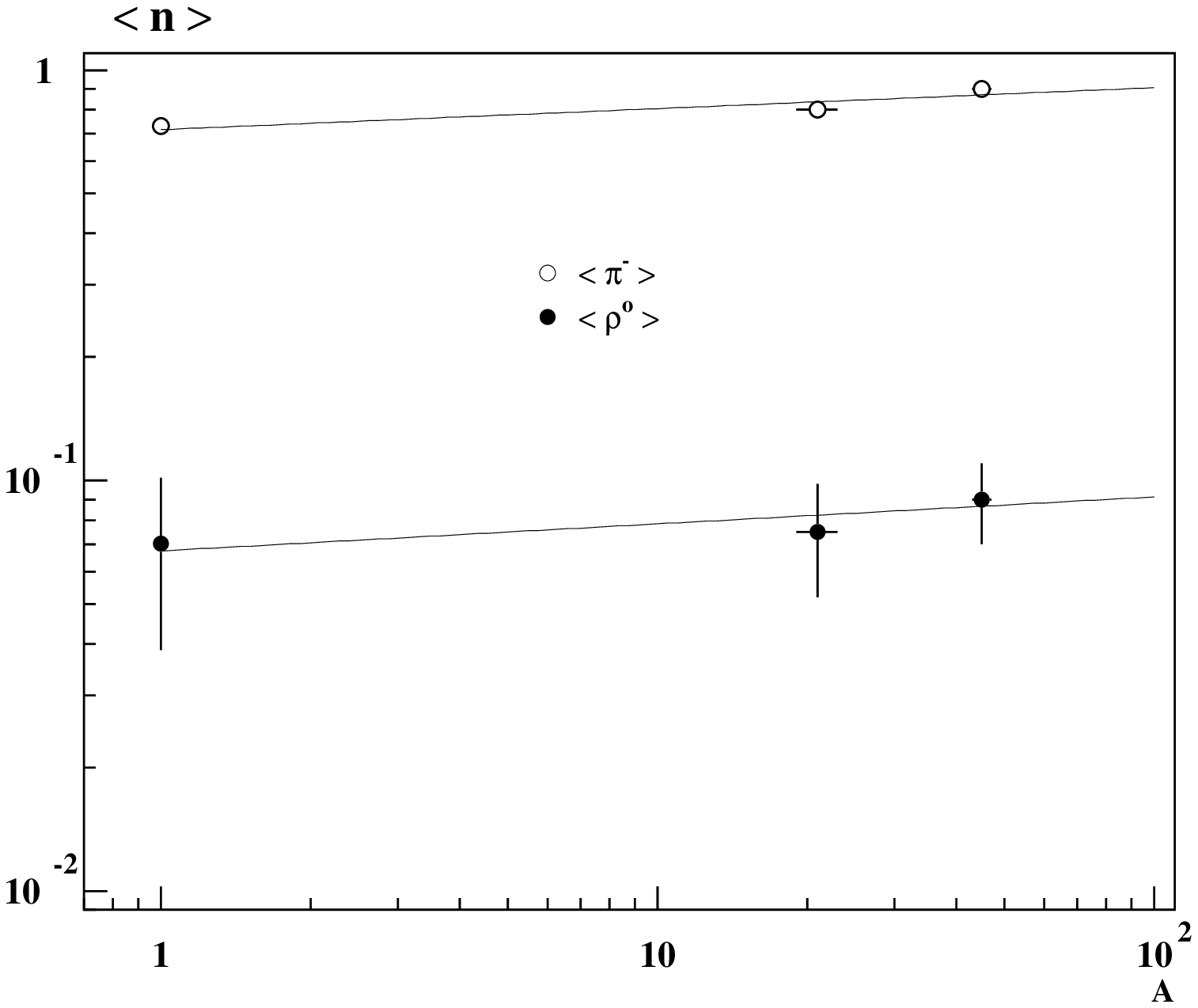}} \caption{The $A$- dependence of the total yields of
$\rho^0$ and $\pi^-$. The lines are the results of the exponential
fit.}
\end{figure}

\newpage
\begin{figure}[h]
\resizebox{1.1 \textwidth}{!}{\includegraphics*[bb=50 120 600
520]{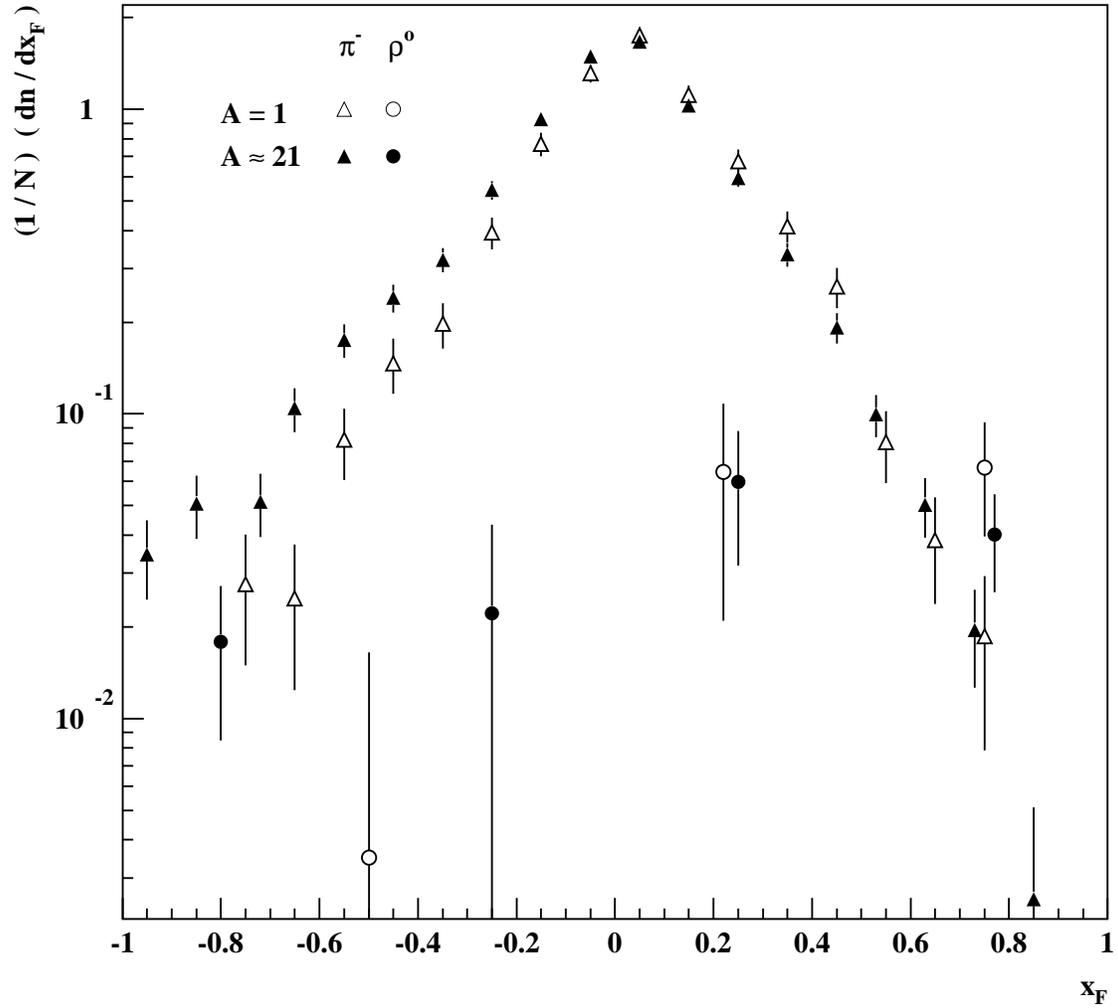}} \caption{The distribution on $x_F$ for $\rho^0$
and $\pi^-$.}
\end{figure}

\newpage
\begin{figure}[ht]
\resizebox{0.9 \textwidth}{!}{\includegraphics*[bb=10 110 600
520]{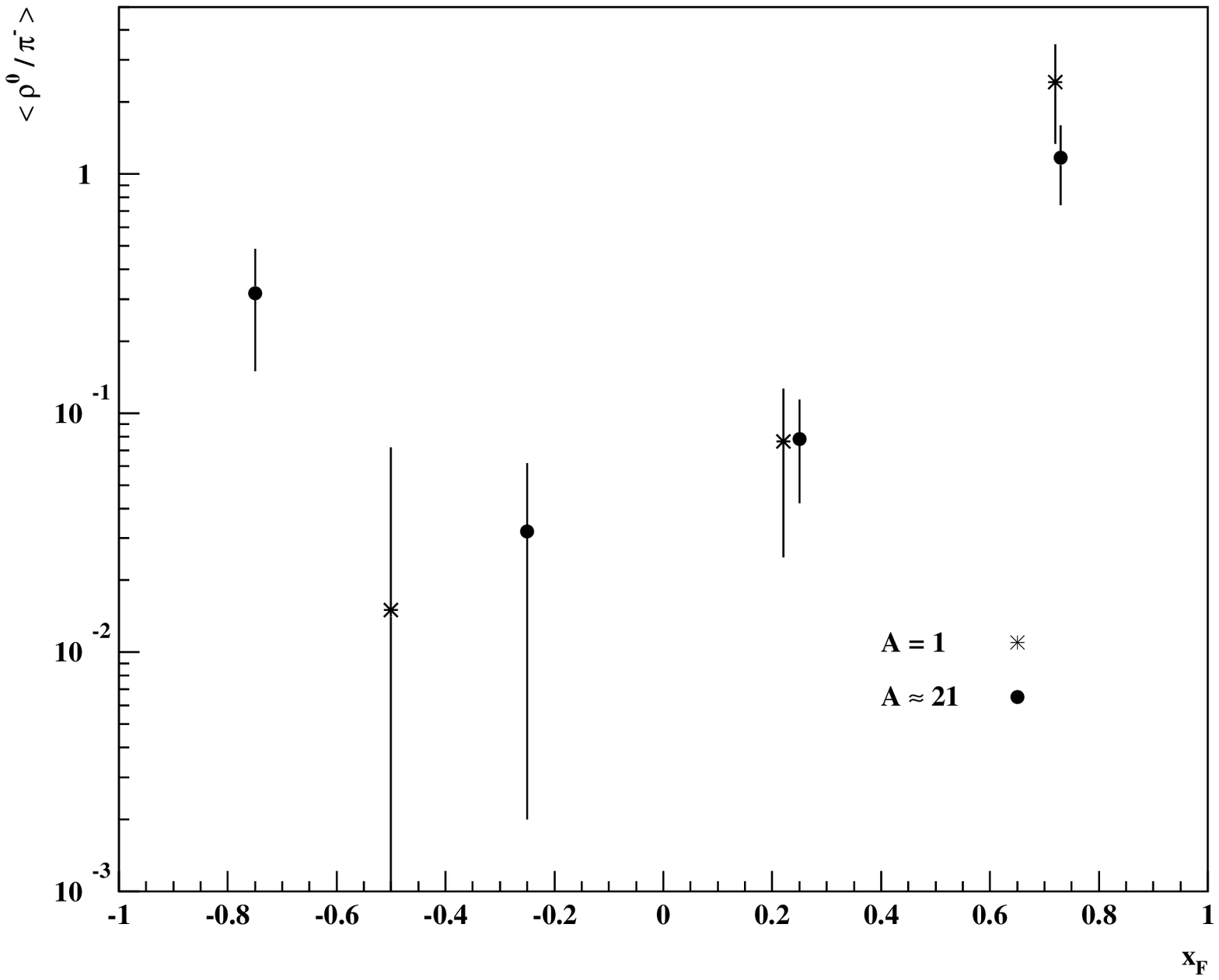}} \caption{The $x_F$- dependence of the ratio
$\rho^0/\pi^-$.}
%\end{figure}
\vspace{0.5cm}
%\newpage
%\begin{figure}[ht]
\resizebox{0.9 \textwidth}{!}{\includegraphics*[bb=50 100 600
520]{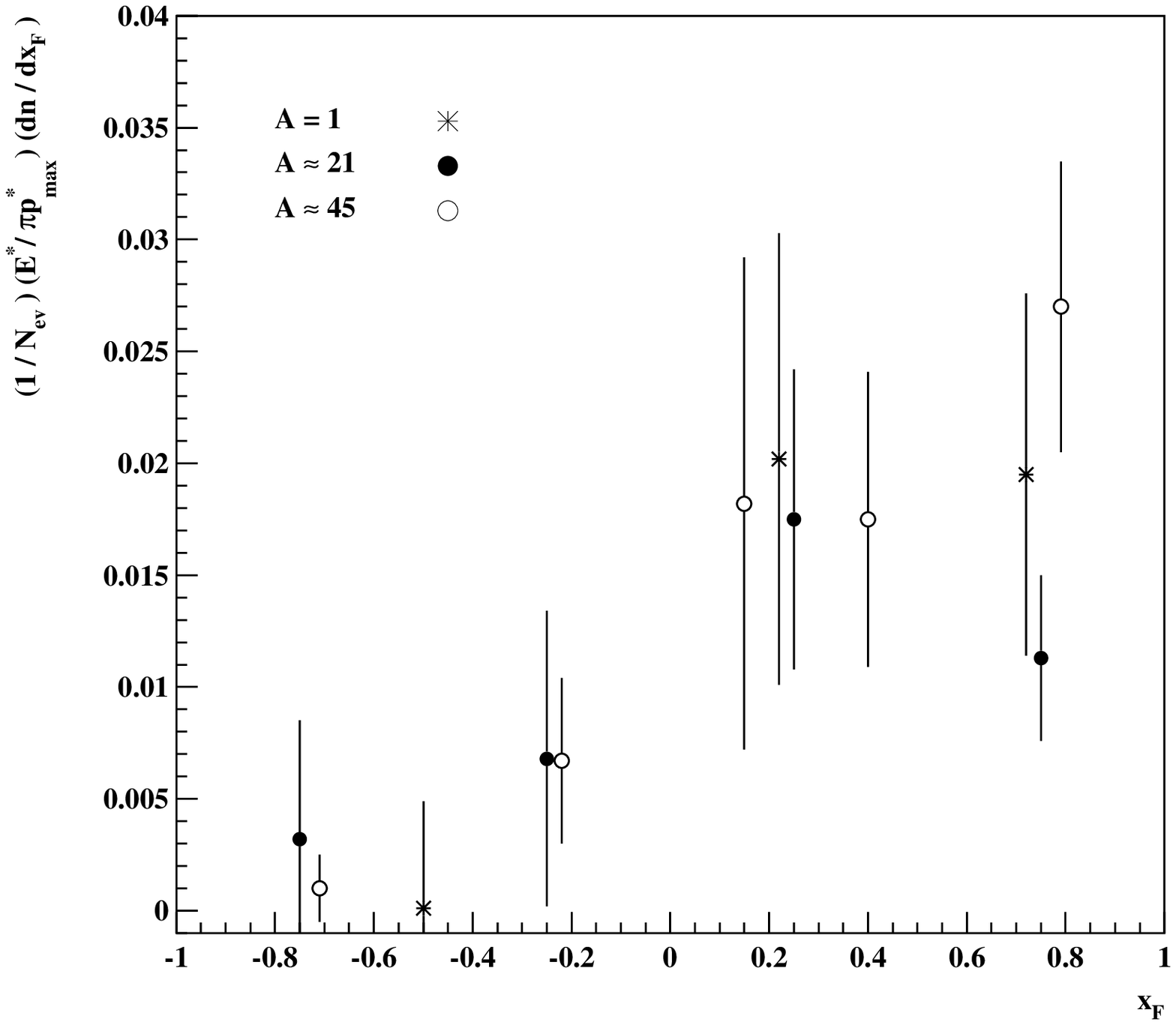}} \caption{The invariant distribution on $x_F$.}
\end{figure}

\newpage
\begin{figure}[ht]
\resizebox{0.9 \textwidth}{!}{\includegraphics*[bb=50 110 600
520]{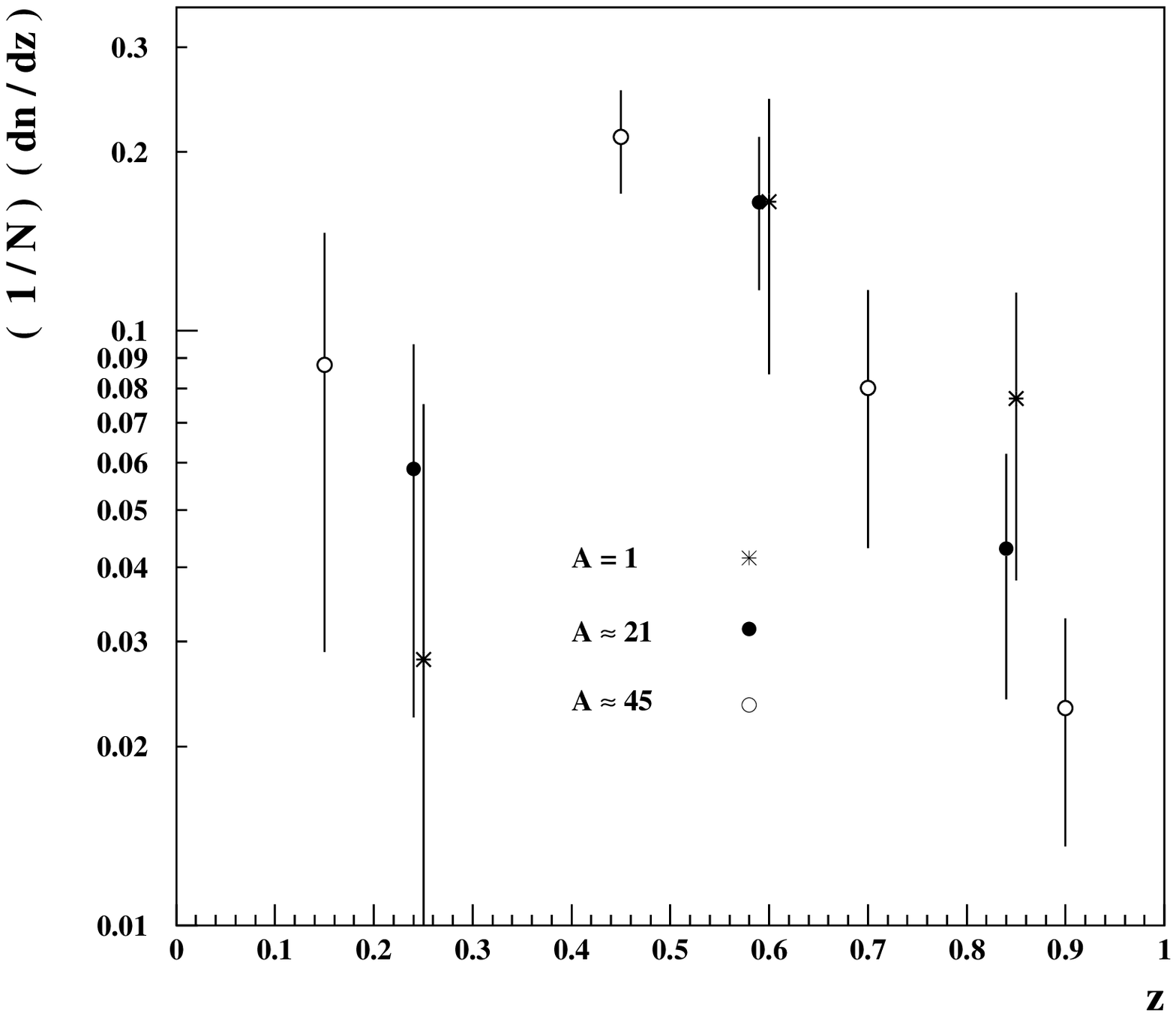}} \caption{The distribution on $z$.}
%\end{figure}
\vspace{0.5cm}
%\newpage
%\begin{figure}[ht]
\resizebox{0.9 \textwidth}{!}{\includegraphics*[bb=50 110 600
520]{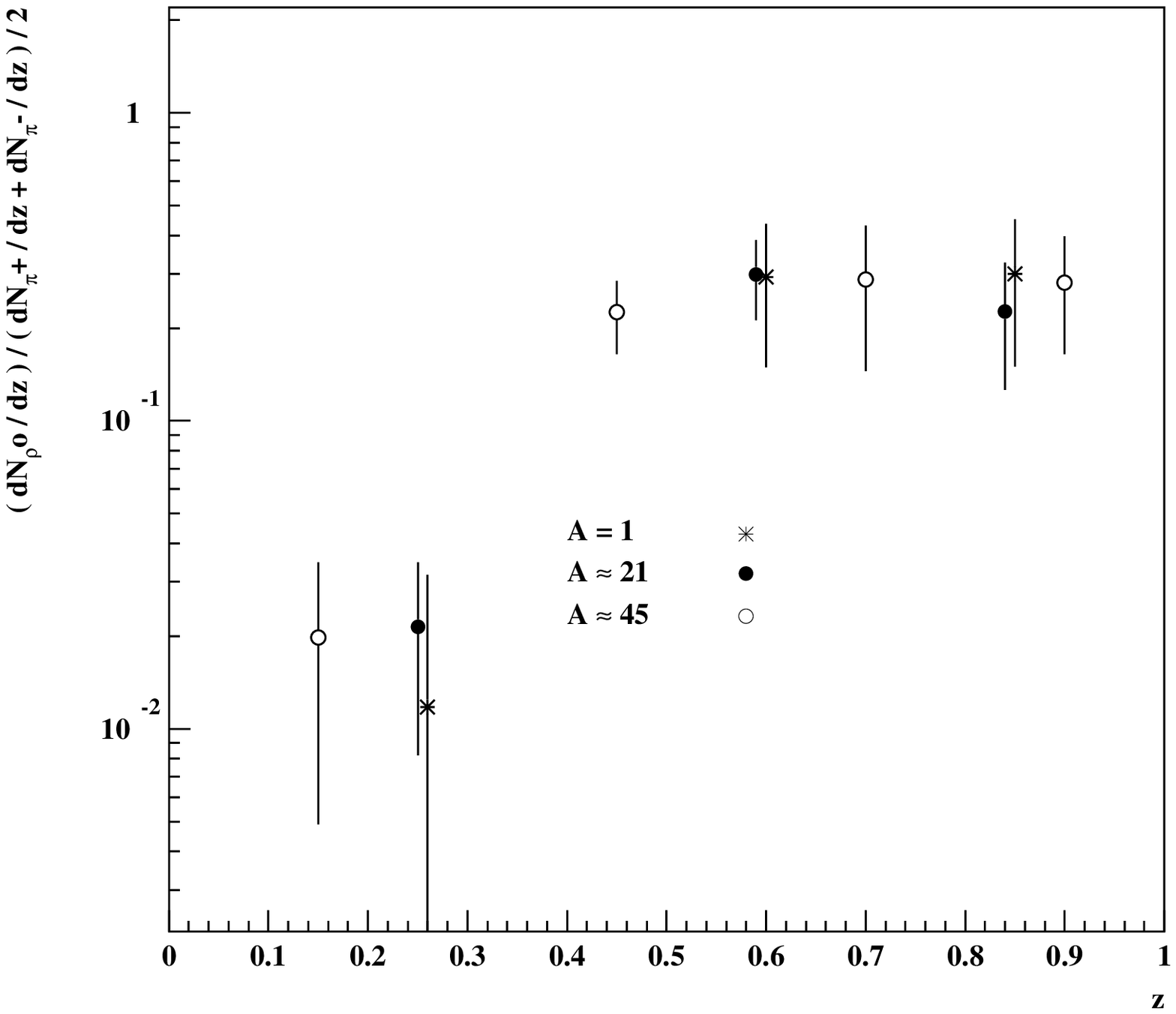}} \caption{The $z$- dependence of the ratio
$\rho^0/({\pi^+} + \pi^-)/2$.}
\end{figure}

%\newpage
%\begin{figure}[ht]
%\resizebox{1.1 \textwidth}{!}{\includegraphics*[bb=50 10 600
%520]{figr8.eps}} \caption{The $p_T^2$- distributions. The lines
%are the result of the exponential fit.}
%\end{figure}


\begin{thebibliography}{00}
\parskip=0.pt \parsep=0.pt \itemsep=0.pt
\bibitem{ref1}
V.V.Ammosov et al., Fiz. Elem. Chastits At. Yadra {\bf23}, 648,
1992 [Sov. J. Part. Nucl. {\bf23}, 283, (1992)].
\bibitem{ref2}
N.M.Agababyan et al., (SKAT Coll), YerPhI Preprint N 1535(9)
(Yerevan, 1999).
\bibitem{ref3}
N.M.Agababyan et al., (SKAT Coll), Yad. Fiz. {\bf66}, 1350 (2003).
[Phys. of At. Nucl. {\bf66}, 1310 (2003)].
\bibitem{ref4}
N.M.Agababyan et al., (SKAT Coll),YerPhI Preprint N 1578(3)
(Yerevan, 2002).
\bibitem{ref5}
N.M.Agababyan et al., (SKAT Coll), Yad. Fiz. {\bf68}, 1241 (2005).
\bibitem{ref6}
N.M.Agababyan et al., (SKAT Coll),YerPhI Preprint N 1597(1)
(Yerevan, 2005); hep-ex/0504024.
\bibitem{ref7}
V.V.Ammosov et al., (SKAT Coll.), Yad. Fiz. {\bf46}, 131 (1987).
[Sov. J. Nucl. Phys. {\bf46}, 80 (1987)].
\bibitem{ref8}
J.D.Jackson, Nuovo Cim. {\bf34}, 1644 (1964).
\bibitem{ref9}
Review of Particle Physics, Phys. Lett. B{\bf592}, 39 (2004).
\bibitem{ref10}
P.Astier et al., (NOMAD Coll.), Nucl. Phys. B{\bf601}, 3 (2001).
\bibitem{ref11}
W.Wittek et al., (BEBC WA59 Coll.), Z. Phys. C{\bf44}, 175 (1989).
\bibitem{ref12}
G.T.Jones et al., (WA21 Coll.), Z. Phys. C{\bf51}, 11 (1991).
\bibitem{ref13}
N.M.Agababyan et al., (SKAT Coll), YerPhI Preprint N 1548(2)
(Yerevan, 2005); hep-ex/0504040.
\bibitem{ref14}
V.Flaminio et al., Compilation CERN-HERA 83-01, 1983.
\end{thebibliography}
\end{document}